\documentclass[aps, prd, twocolumn, nofootinbib,]{revtex4-2}

\usepackage{amsmath}
\usepackage{amsfonts}
\usepackage{wasysym}
\usepackage{graphicx}
\usepackage{comment}
\usepackage{tensor}
\usepackage{cancel}
\usepackage[svgnames]{xcolor}

\usepackage[colorlinks = true, linkcolor = DarkBlue, citecolor = DarkBlue, urlcolor = DarkBlue, bookmarks = false]{hyperref}

\def\filetype{pdf}
\def\path{}

\allowdisplaybreaks[1]

\begin{document}


\title{Rotating dark matter admixed neutron stars}
\author{John Cronin, Xinyang Zhang, and Ben Kain}
\affiliation{Department of Physics, College of the Holy Cross, Worcester, Massachusetts 01610, USA}

\begin{abstract}
\noindent
We study rotating compact stars that are mixtures of the ordinary nuclear matter in a neutron star and fermionic dark matter.  After deriving equations describing a slowly rotating system made up of an arbitrary number of perfect fluids, we specialize to the two-fluid case, where the first fluid describes ordinary matter and the second fluid describes dark matter.  Electromagnetic observations of the moment of inertia and angular momentum directly probe ordinary matter and not dark matter.  Upon taking this into account, we show that the $I$-Love-$Q$ relations for dark matter admixed neutrons stars can deviate significantly from the standard single-fluid relationships.
\end{abstract} 

\maketitle


\section{Introduction}
\label{sec:introduction}

Dark matter admixed neutron stars are mixtures of the ordinary nuclear matter in a neutron star and dark matter modeled as either a bosonic or fermionic particle.  If a sufficient amount of dark matter is present in the star, observable properties such as the mass and radius can be affected.  Models exist in which dark matter is enveloped by ordinary matter and forms a dark matter core or dark matter envelopes ordinary matter and forms a dark matter cloud.  Neutron star observations therefore offer the intriguing possibility of indirectly probing the properties of dark matter.

Static spherically symmetric dark matter admixed neutron stars have been studied extensively.  For a sampling of work in which dark matter is assumed to be a bosonic particle see \cite{Henriques:1989ar, Henriques:1990xg, deSousa:1995ye, deSousa:2000eq, Dzhunushaliev:2011ma, Sanchis-Gual:2022ooi, Shakeri:2022dwg, DiGiovanni:2021ejn} and for dark matter assumed to be a fermionic particle see \cite{Sandin:2008db, Ciarcelluti:2010ji, Leung:2011zz, Leung:2012vea, Goldman:2013qla, Tolos:2015qra, Mukhopadhyay:2015xhs, Gresham:2018rqo, Das:2020vng, Kain:2020zjs, Kain:2021hpk, Das:2021yny, Leung:2022wcf, Hippert:2022snq}.  If one moves away from static and spherically symmetric spacetimes, then there are fewer studies.  Spherically symmetric dynamical studies have simulated dark matter admixed neutron stars with bosonic dark matter \cite{ValdezAlvarado:2012xc, Brito:2015yga, Brito:2015yfh, Bezares:2019jcb, DiGiovanni:2020frc, Valdez-Alvarado:2020vqa, DiGiovanni:2021vlu, Nyhan:2022pda, DiGiovanni:2022xcc} and fermionic dark matter \cite{Gleason:2022eeg}.  Axisymmetric spacetimes allow for rotating dark matter admixed neutron stars.  Reference \cite{DiGiovanni:2022mkn} studied the gravitational waves produced by a rotating star with bosonic dark matter, Ref.~\cite{Guha:2021njn} studied a rotating neutron star with fermionic dark matter using a single-fluid model, and Refs.~\cite{Chan:2021gcm, Chan:2023atg} studied rotating white dwarfs using a Newtonian formalism.  A full three dimensional simulation of a binary inspiral where the stars contained bosonic dark matter was performed in \cite{Bezares:2019jcb}.  Possible formation mechanisms leading to systems with sufficient amounts of dark matter have been considered in \cite{Brito:2015yga, Nelson:2018xtr, Ellis:2018bkr, Deliyergiyev:2019vti, DiGiovanni:2020frc, Gleason:2022eeg, Collier:2022cpr, Chan:2023atg}.

In this work, we study rotating dark matter admixed neutron stars with fermionic dark matter.  As far as we are aware, this has not previously been considered in detail using a multifluid model.  We use Hartle's slowly rotating approximation in which the stationary axisymmetric metric is written as a perturbation about the static spherically symmetric metric with perturbations kept through second order \cite{Hartle:1967he, Hartle:1968si}.  Hartle's formalism requires ordinary matter and dark matter to be rotating uniformly, though it allows them to rotate at different speeds.  By slow, we mean that
\begin{equation} \label{Omega ll}
\Omega_x \ll \sqrt{GM_*/R_*^3},
\end{equation}
where $\Omega_x$ is the angular velocity of either ordinary matter or dark matter, $G$ is the gravitational constant, and $M_*$ and $R_*$ are the mass and radius of the nonrotating star.  

There are a couple important reasons why using the slowly rotating approximation can be more advantageous than numerically solving the full Einstein equations.  First, in the slowly rotating formalism, the angular dependence is described analytically with Legendre polynomials.  This results in a set of linear ordinary differential equations (ODEs) that are significantly easier to solve than the set of nonlinear partial differential equations which follow from the full Einstein equations.  As a consequence, when applicable, the slowly rotating solutions are the most accurate available.  Second, the vast majority of observed neutron stars have rotation speeds consistent with Eq.~(\ref{Omega ll}) \cite{Berti:2004ny}.  The slowly rotating approximation therefore has significant practical value, in addition to being easier to solve.

Dark matter direct detection experiments have placed strong bounds on the interaction strength between nuclear matter and dark matter.  With respect to bulk properties of the star, such as the mass and radius, these bounds indicate that the interaction strength is negligibly small and can be ignored \cite{Nelson:2018xtr, Gresham:2018rqo}.  We model both ordinary matter and fermionic dark matter as perfect fluids.  Our system is then a two-fluid system where the fluids have only gravitational interfluid interactions.

A two-fluid system using Hartle's slowly rotating approximation was developed by Andersson and Comer in \cite{Andersson:2000hy} (see also \cite{Aranguren:2022ted}) for the study of superfluid neutron stars, in which neutrons act as a separate fluid from the remaining charged particles.  Their construction allows for nongravitational interfluid interactions and relies heavily on notation that is largely unfamiliar in the study of dark matter admixed neutron stars.  In this work, we neglect nongravitational interfluid interactions for the reasons explained above and use more familiar notation.  As a generalization to \cite{Andersson:2000hy}, we derive the system of equations for an arbitrary number of perfect fluids.  We then specialize to the two-fluid case, where the first fluid describes ordinary matter and the second fluid describes dark matter.

The use of a perfect fluid to describe matter requires specification of an equation of state, which describes the particle content and interactions between the particles.  Unfortunately, properties of nuclear matter at the extreme pressures found in the core of a neutron star are largely unknown.  As a consequence, the correct equation of state for the ordinary matter in a neutron star is also unknown and many equations of state have been proposed.  The situation is worse for dark matter, since even less is known about the properties of dark matter.  Nonetheless, a choice for the equation of state for dark matter must be made.

The $I$-Love-$Q$ relations for neutron stars are relationships between the moment of inertia $I$, the Love number, and the quadruple moment $Q$ that are universal in that they are approximately independent of the equation of state \cite{Yagi:2013bca, Yagi:2013awa, Yagi:2016bkt, Yeung:2021wvt}.  These relationships were originally discovered using the slowly rotating approximation, but have since been confirmed for rapidly rotating \cite{Pappas:2013naa, Chakrabarti:2013tca} and magnetized \cite{Haskell:2013vha} neutron stars.  The $I$-Love-$Q$ relations allow one to make nontrivial statements about neutron stars without having to rely on a choice for the equation of state.

In this work, we study the $I$-Love-$Q$ relations for dark matter admixed neutron stars.  Two-fluid studies of $I$-Love-$Q$ were recently made for superfluid neutron stars in \cite{Yeung:2021wvt, Aranguren2} using the formalism of \cite{Andersson:2000hy, Aranguren:2022ted}.  We present a complementary two-fluid study of $I$-Love-$Q$ before focusing on dark matter admixed neutron stars.  An important difference between an arbitrary two-fluid system and a dark matter admixed system is that dark matter cannot be probed electromagnetically.  We study how properties of $I$-Love-$Q$ that rely on electromagnetic measurements would then depend on ordinary matter alone, while properties of $I$-Love-$Q$ that depend on gravitational measurements would depend on the full mixed star.  After taking this into account, we find significant deviation from the standard single-fluid $I$-Love-$Q$ relations.  Further, the $I$-Love-$Q$ relations can no longer be parametrized by one-parameter curves.  

This paper is organized as follows.  In Sec.~\ref{sec:equations}, we present a detailed derivation of the system of equations for a slowly rotating star with an arbitrary number of perfect fluids that have only gravitational interfluid interactions.  In Sec.~\ref{sec:numerical}, we explain our numerical methods and present example configurations.  In Sec.~\ref{sec:ILoveQ}, we describe the $I$-Love-$Q$ relationships, present results for a straightforward generalization of $I$-Love-$Q$ for two fluids and results for $I$-Love-$Q$ for dark matter admixed neutron stars.  We conclude in Sec.~\ref{sec:conclusion}.  In writing equations, we use units such that $c = \hbar =1$ and retain the gravitational constant $G$.


\section{Slowly rotating multiple-fluid system}
\label{sec:equations}

In this section, we derive the equations describing an arbitrary number of slowly rotating perfect fluids.  We assume the fluids rotate uniformly about a common rotation axis.  As described in the Introduction, we further assume that the fluids have only gravitational interfluid interactions.

Following Hartle \cite{Hartle:1967he}, we write the stationary axisymmetric metric as a perturbation about a static spherically symmetric metric,
\begin{align}
ds^2 = &-e^{\nu(\bar{r})} \left[ 1 + 2h(\bar{r},\theta) \right] dt^2
\notag 
\\
&+ \frac{\bar{r}}{\bar{r}-2GM(\bar{r})} \left[1 + \frac{2 m(\bar{r},\theta)}{\bar{r} - 2 G M(\bar{r})} \right] d\bar{r}^2
\label{metric}
\\
&+ \bar{r}^2 \left[ 1 + 2k(\bar{r},\theta) \right]
\left\{ d\theta^2 + \sin^2\theta \left[ d\phi - \omega(\bar{r}) dt \right]^2 \right\},
\notag
\end{align}
where
\begin{equation} \label{pert def}
\begin{split}
h(\bar{r},\theta) &= h_0(\bar{r}) + h_2(\bar{r}) P_2(\cos \theta) + O(\Omega^4)
\\
m(\bar{r},\theta) &= m_0(\bar{r}) + m_2(\bar{r}) P_2(\cos \theta) + O(\Omega^4)
\\
k(\bar{r},\theta) &= k_2(\bar{r}) P_2(\cos \theta) + O(\Omega)^4
\\
&= [v_2(\bar{r}) - h_2(\bar{r})] P_2(\cos \theta) + O(\Omega)^4.
\end{split}
\end{equation}
In this metric, $\nu$ and $M$ are equilibrium fields which parametrize the static spherically symmetric metric; $\omega$ is a perturbation that is first order in the angular velocity $\Omega$; and $h$, $h_0$, $h_2$, $m$, $m_0$, $m_2$, $k$, $k_2$, and $v_2$ are second order perturbations.  $P_2(\cos\theta) = (3\cos^2\theta - 1)/2$ is the second order Legendre polynomial.  Subscripts on perturbations indicate whether they are spherical ($\ell = 0$) or quadrupole ($\ell = 2$) perturbations.  The bar on $\bar{r}$ indicates that this is the original radial coordinate.  Soon, we will transform to a new radial coordinate.

The total energy-momentum tensor, $T_\text{tot}^{\mu\nu}$, which is used on the right-hand side of the Einstein field equations,
\begin{equation}
G^{\mu\nu} = 8\pi G T_\text{tot}^{\mu\nu},
\end{equation}
where $G^{\mu\nu}$ is the Einstein tensor, has contributions from each fluid.  With only gravitational interfluid interactions, the total energy-momentum tensor separates and is given by
\begin{equation}
T_\text{tot}^{\mu\nu} = \sum_x T_x^{\mu\nu},
\end{equation}
where $x$ labels the fluid, and
\begin{equation} \label{Tx}
T_x^{\mu\nu}= (\mathcal{E}_x + \mathcal{P}_x)u_x^\mu u_x^\nu + \mathcal{P}_x g^{\mu\nu}
\end{equation}
is the energy-momentum tensor for an individual perfect fluid.  In (\ref{Tx}), $\mathcal{E}_x$ is the energy density, $\mathcal{P}_x$ is the pressure, and $u_x^\mu$ is the four-velocity for fluid $x$.  Each of these quantities depends on $\bar{r}$ and $\theta$ and accounts for effects having to do with rotation.  The four-velocities are given by
\begin{equation}
u_x^\mu = u_x^t(1, 0, 0, \Omega_x),
\end{equation}
where $\Omega_x$ is the angular velocity of the fluid.  Since we are assuming each fluid rotates uniformly, $\Omega_x$ is a constant.  From $g_{\mu\nu} u_x^\mu u_x^\nu = -1$, we have 
\begin{equation} \label{u^t}
u_x^t = \left[ - \left( 
g_{tt} + 2g_{t\phi} \Omega_x + g_{\phi\phi} \Omega_x^2 \right) \right]^{-1/2}.
\end{equation}
As mentioned, the energy density $\mathcal{E}_x$ and the pressure $\mathcal{P}_x$ take into account the effects of rotation.  We can therefore write these as
\begin{equation} \label{eps p perturbations}
\begin{split}
\mathcal{E}_x (\bar{r},\theta)  &= \epsilon_x(\bar{r}) + \delta \mathcal{E}_x (\bar{r},\theta)
\\
\mathcal{P}_x (\bar{r},\theta)  &= p_x(\bar{r}) + \delta \mathcal{P}_x (\bar{r},\theta),
\end{split}
\end{equation}
where $\epsilon_x$ and $p_x$ are the equilibrium fields and $\delta\mathcal{E}_x$ and $\delta \mathcal{P}_{x}$ are perturbations.

With only gravitational interfluid interactions, the equations of state for each fluid also separate,
\begin{equation} \label{barotropic eos}
\mathcal{P}_x = \mathcal{P}_x(\mathcal{E}_x),
\end{equation}
in that $\mathcal{P}_x$ only depends on its respective energy density $\mathcal{E}_x$ and not on properties of other fluids, and the individual energy-momentum tensors are independently conserved,
\begin{equation} \label{cons T}
\nabla_\mu T_x^{\mu\nu} = 0.
\end{equation}


\subsection{Equations of motion}

Conservation of the individual energy-momentum tensors in (\ref{cons T}) lead to the equations of motion
\begin{equation}
\partial_\mu \mathcal{P}_x = (\mathcal{E}_x + \mathcal{P}_x) \partial_\mu ( \ln u^t_x ).
\end{equation}
This equation can be solved analytically.  Defining
\begin{equation} \label{Gamma def}
\Gamma_x \equiv \ln(\mathcal{E}_x + \mathcal{P}_x ) 
- \int_0^{\mathcal{E}_x} \frac{d\mathcal{E}'_x}{\mathcal{E}'_x + \mathcal{P}_x(\mathcal{E}_x')},
\end{equation}
where a prime labels the integration variable, the equations of motion can be written 
\begin{equation} \label{eom0}
\partial_\mu \left( \Gamma_x - \ln u^t_x \right) = 0,
\end{equation}
with solution
\begin{equation} \label{eom}
\Gamma_x - \ln u^t_x  = \ln \mu_x,
\end{equation}
where the $\mu_x$ are constants.

Following \cite{Hartle:1967he, ChandrasekharMiller}, we expand $\Gamma_x$ analogously to the metric fields,
\begin{equation} \label{Gamma expansion}
\Gamma_x(\bar{r},\theta) = \Gamma_x^{\text{eq}}(\bar{r}) + \delta p_{x0}(\bar{r}) + \delta p_{x2}(\bar{r}) P_2(\cos\theta),
\end{equation}
where $ \Gamma_x^{\text{eq}}$ is the equilibrium value and $\delta p_{x0}$ and $\delta p_{x2}$ are second order perturbations.  The use of the symbol $p$ in $\delta p_{x0}$ and $\delta p_{x2}$ is standard, but we note that these are not perturbations to the pressure.  Indeed, the pressure and energy density perturbations are defined in (\ref{eps p perturbations}).  Using that $d \Gamma_x = d\mathcal{P}_x / (\mathcal{E}_x + \mathcal{P}_x)$, one can show that the pressure and energy density perturbations are given by
\begin{align} 
\delta \mathcal{P}_x(\bar{r},\theta) 
&= [\epsilon_x(\bar{r}) + p_x(\bar{r})][\delta p_{x0}(\bar{r}) + \delta p_{x2}(\bar{r}) P_2(\cos\theta)]
\notag
\\
\delta \mathcal{E}_x(\bar{r},\theta) 
&= 
\frac{\partial \epsilon_x(\bar{r})}{\partial p_x(\bar{r})}
\delta \mathcal{P}_x(\bar{r},\theta).
\label{delta eps P}
\end{align}


\subsection{Coordinate transformation}

A subtlety in defining the rotational perturbations arises even at the nonrelativistic level \cite{Hartle:1967he}.  In the absence of rotation, the star is a sphere.  Rotation causes the star to deform and become oblate, pushing matter into regions that previously did not have matter.  The matter in the previously empty regions is described entirely by perturbations, since the equilibrium values vanish.  As a consequence, in these previously empty regions, the perturbations are not small with respect to the equilibrium values.

This can be handled by transforming the radial coordinate \cite{Hartle:1967he},
\begin{equation} \label{xi def}
\bar{r} = r + \xi(r,\theta),
\qquad
\xi(r,\theta) = \xi_0(r) + \xi_2(r) P_2(\cos\theta),
\end{equation}
where $r$ is the new radial coordinate and $\xi$, $\xi_0$, and $\xi_2$ are second order perturbations.\footnote{Hartle writes the original radial coordinate as $r$ and the transformed radial coordinate as $R$ in \cite{Hartle:1967he}.  Our definitions for $\bar{r}$ and $r$ agree with those in \cite{Andersson:2000hy}.}  The edge of the star in both the rotating and nonrotating cases is defined by the smallest radial coordinate such that the total pressure vanishes \cite{Aranguren:2022ted}.  Since we are assuming equations of state that take the barotropic form in (\ref{barotropic eos}), this is equivalent to the smallest radial coordinate such that the total energy density vanishes.  We define the new radial coordinate $r$ such that \cite{Aranguren:2022ted}
\begin{equation} \label{coord transf}
\sum_x \mathcal{P}_x (\bar{r}(r,\theta), \theta) = \sum_x p_x(r).
\end{equation}
This equation can be understood as follows \cite{Hartle:1967he}:~An arbitrary point inside the rotating star resides on a surface of constant total pressure.  The new coordinate $r$ is defined to be the radius of the surface in the nonrotating configuration with the same constant total pressure.

In terms of the new radial coordinate, the energy density and pressure perturbations are well defined.  In transforming these, let $\delta \mathcal{E}_x$ and $\delta \mathcal{P}_x$ be the perturbations in the original coordinate system and let $\Delta \mathcal{E}_x$ and $\Delta \mathcal{P}_x$ be the perturbations in the new coordinate system.  The transformations are then
\begin{equation} \label{delta script P R} 
\begin{split}
\Delta \mathcal{E}_x(r,\theta) &= 
\delta \mathcal{E}_x(r,\theta) 
+ \xi(r,\theta) \frac{d \epsilon_x(r)}{d r}
\\
\Delta \mathcal{P}_x(r,\theta) &= 
\delta \mathcal{P}_x(r,\theta) 
+ \xi(r,\theta) \frac{d p_x(r)}{d r}.
\end{split}
\end{equation}
These may be derived by inserting (\ref{xi def}) into (\ref{eps p perturbations}) and then expanding through second order in perturbations.  This same calculation in (\ref{coord transf}) leads to
\begin{equation} \label{sum Delta P}
\sum_x \Delta \mathcal{P}_x(r,\theta) = 0.
\end{equation}

Once solutions are found, we can return to the original radial coordinate.  To do so, we will need expressions for $\xi_0$ and $\xi_2$ in (\ref{xi def}).  We can obtain such expressions from the transformations for the pressure perturbations.  Using (\ref{sum Delta P}) in (\ref{delta script P R}) gives
\begin{equation} \label{sum delta eps P}
\sum_x \delta \mathcal{P}_x(r,\theta) = - \xi(r,\theta) 
\frac{d}{dr}\sum_x 
p_x(r).
\end{equation}
Combining Eqs.~(\ref{delta eps P}), (\ref{xi def}), and (\ref{sum delta eps P}), we find the desired formulas for the radial coordinate perturbations,
\begin{equation} \label{xi eq}
\begin{split}
\xi_0(r) 
&=
- \frac{\sum_x [\epsilon_x(r) + p_x(r)] \delta p_{x0}(r)}
{\sum_y d p_y(r)/d r} 
\\
\xi_2(r) 
&=
- \frac{\sum_x [\epsilon_x(r) + p_x(r)] \delta p_{x2}(r)}
{\sum_y d p_y(r)/d r}.
\end{split}
\end{equation}

We must write all equations in terms of the new radial coordinate, since it is the new radial coordinate that leads to well-defined energy density and pressure perturbations.  This requires transforming the Einstein tensor and the energy-momentum tensor.  Let $\delta G^{\mu\nu}$ and $\delta T_x^{\mu\nu}$ be the perturbations in the original coordinate system and let $\Delta G^{\mu\nu}$ and $\Delta T_x^{\mu\nu}$ be the perturbations in the new coordinate system.  The coordinate transformations are then
\begin{equation}  \label{coord transformation equation}
\begin{split}
\Delta G^{\mu\nu}(r,\theta) 
= \delta G^{\mu\nu}(r,\theta) 
+ \xi(r,\theta) \frac{d G^{\mu\nu}(r)}{d r}
\\
\Delta T_x^{\mu\nu}(r,\theta) 
= \delta T_x^{\mu\nu}(r,\theta) 
+ \xi(r,\theta) \frac{d T_x^{\mu\nu}(r)}{d r},
\end{split}
\end{equation}
where $G^{\mu\nu}(r)$ and $T_x^{\mu\nu}(r)$ in the derivatives are equilibrium values.

The equilibrium energy-momentum tensor for an individual fluid is
\begin{equation}
[T_x(r)]\indices{^\mu_\nu} = \text{diag} [-\epsilon_x(r), p_x(r), p_x(r),p_x(r)].
\end{equation}
Using this and the equilibrium Einstein field equations, the components of the Einstein tensor perturbations we will need are
\begin{align} 
\Delta G\indices{^t_t}(r,\theta) 
&= \delta G\indices{^t_t}(r,\theta) 
- 8\pi G \xi(r,\theta) \frac{d}{dr} \sum_x\epsilon_x (r)
\notag
\\
\Delta G\indices{^r_r}(r,\theta) 
&= \delta G\indices{^r_r}(r,\theta) 
+ 8\pi G \xi(r,\theta) \frac{d}{dr} \sum_x p_x (r)
\notag
\\
\Delta G\indices{^\theta_\theta}(r,\theta) 
&= \delta G\indices{^\theta_\theta}(r,\theta) 
+ 8\pi G \xi(r,\theta) \frac{d}{dr} \sum_x p_x (r)
\notag
\\
\Delta G\indices{^\phi_\phi}(r,\theta) 
&= \delta G\indices{^\phi_\phi}(r,\theta) 
+ 8\pi G \xi(r,\theta) \frac{d}{dr} \sum_x p_x (r)
\notag
\\
\Delta G\indices{^r_\theta}(r,\theta) 
&= \delta G\indices{^r_\theta}(r,\theta)
\notag
\\
\Delta G\indices{^t_\phi}(r,\theta) 
&= \delta G\indices{^t_\phi}(r,\theta).
\label{Einstein coord transf}
\end{align}

For the components of the energy-momentum tensor perturbations, we plug the energy density and pressure perturbations in (\ref{eps p perturbations}) into (\ref{Tx}) and use (\ref{u^t}).  Expanding the results through second order in perturbations and canceling terms using (\ref{sum Delta P}), we find
\begin{align}
[\Delta T_\text{tot}(r,\theta)]\indices{^t_t} 
&= - \frac{r^2\sin^2\theta}{e^{\nu(r)}}
\sum_x 
[\epsilon_x(r) + p_x(r)]
\varpi_x(r)
\Omega_x
\notag
\\
&\qquad
- \sum_x \Delta \mathcal{E}_x (r,\theta)
\notag
\\
[\Delta T_\text{tot}(r,\theta)]\indices{^\phi_\phi} 
&= + \frac{r^2\sin^2\theta}{e^{\nu(r)}}
\sum_x 
[\epsilon_x(r) + p_x(r)]
\varpi_x(r)
\Omega_x
\notag
\\
[\Delta T_\text{tot}(r,\theta)]\indices{^t_\phi} 
&= + \frac{r^2\sin^2\theta}{e^{\nu(r)}} 
\sum_x 
[\epsilon_x(r) + p_x(r)]
\varpi_x(r),
\label{EM comps}
\end{align}
where
\begin{equation} \label{varpi def}
\varpi_x(r) \equiv \Omega_x - \omega(r),
\end{equation}
with all other components vanishing.  In particular,
\begin{equation} \label{EM comps zero}
(\Delta T_\text{tot})\indices{^r_r} = 0,
\qquad
(\Delta T_\text{tot})\indices{^\theta_\theta} = 0,
\qquad
(\Delta T_\text{tot})\indices{^r_\phi} = 0,
\end{equation}
which we shall use.  In (\ref{EM comps}), $\sin^2\theta$ can be written as
\begin{equation} \label{sin decomp}
\sin^2\theta = \frac{2}{3} - \frac{2}{3} P_2(\cos\theta).
\end{equation}


\subsection{Integrals of motion}

The equations of motion in (\ref{eom}) are written in terms of the constants $\mu_x$.  Following Hartle \cite{Hartle:1967he}, we can write these constants as
\begin{equation} \label{mu_x expansions}
\mu_x = \mu_x^\text{eq} \left[ 1 + \gamma_x + O(\Omega^4) \right],
\end{equation}
where $\mu_x^\text{eq}$ are the equilibrium values and the $\gamma_x$ are constants that are second order in $\Omega$.  To derive the integrals of motion, we insert the expansion for $\Gamma_x$ in (\ref{Gamma expansion}), $u^t_x$ in (\ref{u^t}), and  $\mu_x$ in (\ref{mu_x expansions}) into the equations of motion and then expand through second order in the perturbations.  We then transform to the new radial coordinate and use (\ref{sin decomp}).  We find for the integrals of motion
\begin{equation} \label{integrals of motion}
\begin{split}
\gamma_x &= \delta p_{x0}(r) + h_0(r)
- \frac{1}{3} r^2  e^{-\nu(r)} \varpi^2_x(r)
\\
0 &= \delta p_{x2}(r) + h_2(r)
+ \frac{1}{3} r^2  e^{-\nu(r)} \varpi^2_x(r),
\end{split}
\end{equation}
where $\varpi_x$ is defined in (\ref{varpi def}).


\subsection{Einstein field equations}

In this subsection, we present the Einstein field equations, which are the principle equations we solve in our study of rotating dark matter admixed neutron stars.  The Einstein tensor for stationary axisymmetric spacetimes can be found, for example, in \cite{Hartle:1967he, ChandrasekharFriedman}.  In the exterior of the star, all fluids have vanishing energy density and pressure.  As a consequence, the solutions to the Einstein field equations take the same form in the exterior as they do in the single-fluid case and we can use the analytical exterior solutions presented in \cite{Hartle:1967he}.


\subsubsection{Equilibrium:~$\nu$ and $M$}

For the equilibrium equations, we drop all perturbations.  The Einstein field equations and the equations of motion reduce to the multifluid Tolman–Oppenheimer–Volkoff (TOV) equations,
\begin{equation} \label{TOV}
\begin{split}
\frac{d\nu}{dr} &= 
\frac{8\pi G r^3 \sum_x p_x  + 2GM}{r(r-2GM)}
\\
\frac{dM}{dr} &= 4\pi r^2 \sum_x \epsilon_x
\\
\frac{dp_x}{dr} &= - \frac{1}{2} (\epsilon_x + p_x) \frac{d\nu}{dr}.
\end{split}
\end{equation}
These equations may be integrated outward from $r = 0$ using the inner boundary conditions 
\begin{align}
\nu(r) &= \nu(0) + r^2\frac{4\pi G}{3} \sum_x [\epsilon_x(0) + 3p_x(0)] + O(r^4)
\notag 
\\
M(r) &= r^3 \frac{4\pi}{3} \sum_x \epsilon_x(0) + O(r^5)
\notag
\\
p_x(r) &= p_x(0) - r^2 \frac{2\pi G}{3} [\epsilon_x(0) + p_x(0)] \sum_y [\epsilon_y(0) + 3p_y(0)]
\notag
\\
&\qquad + O(r^4).
\end{align}
The edge of each fluid occurs at the smallest value of $r = R_x$, such that
\begin{equation}
p_x(R_x) = 0.
\end{equation}
We label the edge of the outermost fluid as $r = R_*$, which marks the edge of the star.  The total nonrotating mass of the star is given by
\begin{equation} \label{M* def}
M_* = M(R_*).
\end{equation}

In the exterior of the star, the equilibrium solution is the Schwarzschild solution.  The interior and exterior solutions must match at $r = R_*$, which means
\begin{equation} \label{nu ext}
e^{\nu(R_*)} = 1 - \frac{2 G M_*}{R_*}
\end{equation}
and that $M_*$ gives the Arnowitt-Deser-Misner mass.


\subsubsection{First order:~$\omega$}

The first order equation follows from the $t\phi$ component.  The $t\phi$ component of the Einstein tensor is
\begin{equation}
G\indices{^t_\phi} = - \frac{\sin^2\theta}{2\bar{r}^2} j
\frac{d}{d\bar{r}} \left( \bar{r}^4 j \frac{d\omega}{d\bar{r}} \right),
\end{equation}
where
\begin{equation}
j \equiv e^{-\nu/2} \sqrt{1 - 2GM/\bar{r}}.
\end{equation}
Using (\ref{Einstein coord transf}) and (\ref{EM comps}) we find for the Einstein field equation
\begin{equation} \label{1st order eq}
\sum_x (\epsilon_x + p_x)
\left[ - \frac{1}{r^4} \frac{d}{dr} \left(  r^4  j \frac{ d \omega}{dr} \right)
+ \frac{4}{r} \frac{dj}{dr} \varpi_x \right] = 0.
\end{equation}

Equation (\ref{1st order eq}) is a second order ODE.  To facilitate solving it numerically, we write it as a system of first order ODEs.  Defining
\begin{equation}
u(r) \equiv - r^4 j(r) \frac{d\omega(r)}{dr},
\qquad
\eta_x(r) \equiv j(r)  \varpi_x(r),
\end{equation}
Eq.~(\ref{1st order eq}) is equivalent to
\begin{equation} \label{chi u eqs}
\begin{split}
\frac{du}{dr} &= \frac{16\pi G r^5}{r - 2GM} 
\sum_x  (\epsilon_x + p_x) \eta_x.
\\
\frac{d\eta_x}{dr} &= \frac{u}{r^4} 
- \frac{4\pi G r^2 \eta_x}{r-2GM} \sum_y \left(\epsilon_y  + p_y \right).
\end{split}
\end{equation}
Notice that each term on the right-hand side of the $du/dr$ equation is for an individual fluid.  It will prove useful to decompose $u$ as
\begin{equation} \label{u decom}
u(r) \equiv \sum_x u_x(r),
\end{equation}
where the $u_x$ satisfy
\begin{equation} \label{dux dr}
\frac{du_x}{dr} = \frac{16\pi G r^5}{r - 2GM} 
(\epsilon_x + p_x) \eta_x.
\end{equation}
These equations may be integrated outward from $r = 0$ using the inner boundary conditions
\begin{align}
u_x(r) &= r^5 \frac{16\pi G}{5} [\epsilon_x(0) + p_x(0)] \eta_x(0) + O(r^7)
\notag
\\
\eta_x(r) &= \eta_x(0) 
+ r^2 \biggl\{
\frac{8\pi G}{5} \sum_y [\epsilon_y(0) + p_y(0)]\eta_y(0)
\notag
\\
&\qquad
- 2\pi G \eta_x(0)\sum_y [\epsilon_y(0) + p_y(0)] 
\biggr\}
+ O(r^4),
\end{align}
where
\begin{equation} \label{eta_x(0)}
\eta_x(0) =e^{-\nu(0)/2} \varpi_x(0).
\end{equation}

The interior and exterior solutions must match at $r = R_*$, which leads to \cite{Hartle:1967he}
\begin{equation} \label{u eta ext}
u(R_*) = 6J, \qquad
\eta_x(R_*) = \Omega_x - \frac{2J}{R_*^3},
\end{equation}
where $J$ is the total angular momentum of the system.  We expect the total angular momentum to be equal to the sum of the angular momentum for each fluid.  Using the decomposition of $u$ in (\ref{u decom}), we can see that \cite{Leung:2012vea}
\begin{equation}
u_x(R_*) = 6J_x,
\end{equation}
where
\begin{equation}
J = \sum_x J_x.
\end{equation}
With these equations we can find the angular momentum and angular velocity of each fluid.  We can then compute the moment of inertia for each fluid,
\begin{equation} \label{Ix def}
I_x = \frac{J_x}{\Omega_x},
\end{equation}
and the total moment of inertia of the system,
\begin{equation}
I = \sum_x I_x.
\end{equation}


\subsubsection{Second order:~$m_0$ and $\delta p_{x0}$}

The relevant Einstein tensor perturbations are
\begin{align} 
(\delta G\indices{^t_t})_{\ell = 0}
&= \frac{j\omega }{3\bar{r}^2}
\frac{d}{d\bar{r}} 
\left(\bar{r}^4  j \frac{d\omega}{d\bar{r}} \right)
+ \frac{1}{6} \bar{r}^2 j^2 \left(\frac{d\omega}{d\bar{r}} \right)^2 - \frac{2}{\bar{r}^2} \frac{d m_0}{d\bar{r}} 
\notag
\\
(\delta G\indices{^r_r})_{\ell = 0}
&= 
\frac{1}{6} \bar{r}^2
j^2
\left(\frac{d\omega}{d\bar{r}}\right)^2
- \left( \frac{d\nu}{d\bar{r}} + \frac{1}{\bar{r}} \right) 
\frac{2m_0}{\bar{r}^2}
\notag
\\
&\qquad
+ \left(1 - \frac{2GM}{\bar{r}} \right)
\frac{2}{\bar{r}}
\frac{dh_0}{d\bar{r}}.
\end{align}
To derive the desired equations, we use (\ref{Einstein coord transf}), (\ref{EM comps}), and (\ref{EM comps zero}).  We also use (\ref{delta script P R}), (\ref{xi eq}), (\ref{1st order eq}), the derivative of the first integral of motion in (\ref{integrals of motion}), and the $d\nu/dr$ TOV equation in (\ref{TOV}).  We find
\begin{align} 
\frac{d m_0}{dr} 
&= 
\frac{u^2}{12 r^4}
+ \frac{8\pi G r^5}{3(r - 2GM)}  \sum_x (\epsilon_x + p_x) \eta_x^2
\notag
\\
&\qquad
+ 4\pi G r^2 \sum_x \frac{\partial \epsilon_x}{\partial p_x} \delta p_{x0}(\epsilon_x + p_x)
\notag
\\
\frac{d\delta p_{x0}}{dr} 
&=
\frac{u^2}{12 r^4 (r-2GM)}
- \frac{m_0(1 + 8\pi G r^2 \sum_y p_y )} {(r-2GM)^2}
\notag
\\
&\qquad
- \frac{4\pi G r^2}{r-2GM} \sum_y(\epsilon_y + p_y) \delta p_{y0}
\notag
\\
&\qquad
+ \frac{2 r^2 \eta_x}{3(r-2GM)}
\biggl[
\frac{u}{r^3}
\notag
\\
&\qquad\qquad
+  \frac{\eta_x (r - 3 GM - 4\pi G r^3 \sum_y p_y)}{r-2GM}
\biggr].
\label{m0 deltapx0 eqs}
\end{align}
These equations may be integrated outward from $r = 0$ using the inner boundary conditions
\begin{align}
m_0(r) &= r^5 \frac{4\pi G}{15} \sum_x \left( \frac{d\epsilon_x}{dp_x} \biggr|_{r=0} + 2\right) 
\notag
\\
&\qquad \times[\epsilon_x(0) + p_x(0) ] \eta_x^2(0)
+ O(r^7)
\notag
\\
\delta p_{x0}(r) &=  r^2 \frac{\eta_x^2(0)}{3} + O(r^4).
\end{align}

The interior and exterior solutions must match at $r = R_*$, which leads to \cite{Hartle:1967he}
\begin{equation} \label{delta M}
m_0(R_*) = \delta M -\frac{J^2}{R_*^3},
\end{equation}
where $\delta M$ is the correction to the mass of the star due to rotation.  The total mass of the star is given by
\begin{equation}
\mathcal{M} = M_* + \delta M.
\end{equation}


\subsubsection{Second order:~$m_2$}

The relevant Einstein tensor perturbations are
\begin{align} 
\delta G\indices{^\theta_\theta} - \delta  G\indices{^\phi_\phi}
&= \sin^2\theta 
\biggl[ - \frac{3}{\bar{r}^2} \left( h_2 + \frac{m_2}{\bar{r}- 2GM} \right) 
\notag
\\
&\qquad
+ \frac{1}{2} j^2 \bar{r}^2 
\left( \frac{d\omega}{d\bar{r}}\right)^2 
+ \frac{j \omega}{2\bar{r}^2}
\frac{d}{d\bar{r}} 
\left(
\bar{r}^4  j
\frac{d\omega}{d\bar{r}} \right) \biggr].
\end{align}
Using (\ref{Einstein coord transf}), (\ref{EM comps}), and (\ref{EM comps zero}), we find the algebraic equation
\begin{equation} \label{m2 eq}
\frac{m_2}{r-2GM} = 
-h_2
+ \frac{u^2}{6}
+ \frac{8\pi G}{3} r^4 e^{-\nu}\sum_x \varpi_x^2 (\epsilon_x + p_x).
\end{equation}


\begin{widetext}

\subsubsection{Second order:~$v_2$ and $h_2$}

The relevant Einstein tensors are
\begin{equation} 
\begin{split}
(\delta G\indices{^r_r})_{\ell = 2}
&= 
\frac{1}{6} \bar{r}^2
j^2
\left(\frac{d \omega}{d\bar{r}}\right)^2
- \left( \frac{d\nu}{d\bar{r}} + \frac{1}{\bar{r}} \right) 
\frac{2m_2}{\bar{r}^2}
+ \left(1 - \frac{2GM}{\bar{r}} \right)
\left[\frac{2}{\bar{r}} \frac{d h_2}{d\bar{r}}
+ \frac{d k_2}{d\bar{r}} \left( \frac{d\nu}{d\bar{r}} + \frac{2}{\bar{r}} \right) \right]
-\frac{1}{\bar{r}^2} 
(6 h_2 + 4k_2)
\\
(\delta G\indices{^r_\theta})_{\ell = 2}
&= - \frac{dh_2}{d\bar{r}} + h_2 \left( \frac{1}{\bar{r}} - \frac{1}{2}\frac{d\nu}{d\bar{r}} \right) - \frac{dk_2}{d\bar{r}} + \frac{m_2}{\bar{r}-2GM} \left( \frac{1}{\bar{r}} + \frac{1}{2}\frac{d\nu}{d\bar{r}} \right).
\end{split}
\end{equation}
To derive the desired equations, we use (\ref{Einstein coord transf}), (\ref{EM comps}), and (\ref{EM comps zero}).  We also use (\ref{xi eq}), the second integral of motion in (\ref{integrals of motion}), (\ref{m2 eq}), and write $k_2 = v_2 - h_2$ from (\ref{pert def}).  We find
\begin{equation} \label{v2 h2 eqs}
\begin{split}
\frac{dv_2}{dr} 
&= -\nu' h_2
+  \left( \frac{1}{r} + \frac{\nu'}{2} \right)
\left[ \frac{u^2}{6 r^4}
+ \frac{8\pi G r^5 }{3(r-2GM)}\sum_x \eta_x^2 (\epsilon_x + p_x) \right]
\\
\frac{h_2}{dr}
&= h_2 \left\{ - \nu' + \frac{r}{r-2GM} \frac{1}{\nu'} \left[ 8\pi G \sum_x (\epsilon_x + p_x) - \frac{4 GM}{r^3} \right] \right\}
-\frac{4 v_2}{r(r-2GM)} \frac{1}{\nu'} 
\\
&\qquad
+ \frac{u^2}{6 r^5}
\left( \frac{r\nu'}{2} 
- \frac{1}{r-2GM} \frac{1}{\nu'} \right)
+
\frac{8\pi G  r^4}{3(r - 2MG)}
\left( \frac{r\nu'}{2} 
+ \frac{1}{r-2GM} \frac{1}{\nu'} \right)
\sum_x \eta_x^2 (\epsilon_x + p_x),
\end{split}
\end{equation}
\end{widetext}
where $\nu' \equiv d\nu/dr$ is given by the TOV equation in (\ref{TOV}).  The inner boundary conditions for these equations are
\begin{equation}
h_2(r) = A r^2 + O(r^4),
\qquad
v_2(r) = B r^4 + O(r^6),
\end{equation}
where
\begin{equation} \label{A B formula}
\begin{split}
B 
&= \frac{2\pi G}{3} \sum_x [\epsilon_x(0) + p_x(0)]\eta_x^2(0)
\\
&\qquad
-
A \frac{2\pi G}{3} \sum_x[\epsilon_x(0) + 3 p_x(0) ].
\end{split}
\end{equation}
The value for $A$ is determined by an outer boundary condition \cite{Hartle:1967he}.  

Following \cite{Hartle:1967he} we can construct the general solution by writing it as the sum of particular and complementary solutions,
\begin{equation} \label{interior v2 h2}
v_2(r) = v_2^P(r) + C v_2^C(r),
\qquad
h_2(r) = h_2^P(r) + C h_2^C(r),
\end{equation}
where $C$ is a constant.  The particular solution is \textit{any} solution to (\ref{v2 h2 eqs}), which means we may choose the value of $A$ arbitrarily (in practice, we use $A=1$) and then choose $B$ according to (\ref{A B formula}).  The complementary solution is the solution to the complementary equations, which are the homogenous versions of (\ref{v2 h2 eqs}) and are obtained by setting $u = \eta_x = 0$ in (\ref{v2 h2 eqs}),
\begin{equation} \label{v2 h2 eqs hom}
\begin{split}
\frac{dv_2^C}{dr} 
&= -\nu' h_2^C
\\
\frac{h_2^C}{dr}
&= h_2^C \biggl\{ - \nu' + \frac{r}{r-2GM} \frac{1}{\nu'} 
\\
&\qquad 
\times
\left[ 8\pi G \sum_x (\epsilon_x + p_x) - \frac{4 GM}{r^3} \right] \biggr\}
\\
&\qquad
-\frac{4 v_2^C}{r(r-2GM)} \frac{1}{\nu'}.
\end{split}
\end{equation}
The inner boundary conditions for the complementary equations are
\begin{equation}
h_2^C(r) = a r^2 + O(r^4),
\qquad
v_2^C(r) = b r^4 + O(r^6),
\end{equation}
where
\begin{equation} \label{a b formula}
b = - a \frac{2\pi G}{3} \sum_x[\epsilon_x(0) + 3 p_x(0) ].
\end{equation}
$a$ may be chosen arbitrarily (in practice, we use $a = 1$) with $b$ chosen according to (\ref{a b formula}), since the integration constant is accounted for by $C$ in (\ref{interior v2 h2}).  Equation (\ref{interior v2 h2}) gives the general solution when $C$ is arbitrary.

The interior solution in (\ref{interior v2 h2}) must match the exterior solution at $r = R_*$.  This leads to \cite{Hartle:1967he}
\begin{equation} \label{v2 h2 exterior}
\begin{split}
 v_2^P(R_*) + C v_2^C(R_*) &= - \frac{J^2}{R_*^4} + K \frac{2 M_*}{\sqrt{R_*(R_*-2M_*)}} Q^2_1(\zeta)
\\
 h_2^P(R_*) + C h_2^C(R_*) &= J^2 \left( \frac{1}{M_* R_*^3} + \frac{1}{R_*^4} \right) + K Q_2^2(\zeta),
\end{split}
\end{equation}
where $K$ is a constant, 
\begin{equation}
\begin{split}
Q_1^2(\zeta) &= (\zeta^2 - 1)^{1/2} \left[ \frac{3\zeta^2 - 2}{\zeta^2-1} - \frac{3}{2} \zeta\ln\left( \frac{\zeta+1}{\zeta-1} \right) \right]
\\
Q_2^2(\zeta) &= \frac{3}{2} (\zeta^2 - 1) \ln\left(\frac{\zeta + 1}{\zeta-1}\right) - \frac{3\zeta^3 - 5\zeta}{\zeta^2 -1} 
\end{split}
\end{equation}
are associated Legendre functions of the second kind, and
$\zeta = R_*/M_* - 1$.  From (\ref{v2 h2 exterior}), we can solve for $C$ and $K$.  Using $C$ in (\ref{interior v2 h2}), we have the interior solution.  With $K$, we can compute the quadrupole moment \cite{Hartle:1968si},
\begin{equation} \label{Q def}
Q = \frac{J^2}{M_*} + \frac{8}{5} K M_*^3.
\end{equation}


\subsubsection{Second order:~$\delta p_{x2}$}

$\delta p_{x2}$ is found algebraically using the second integral of motion in (\ref{integrals of motion}).  Having found $\delta p_{x0}$ and $\delta p_{x2}$, we can compute the radial coordinate perturbations $\xi_0$ and $\xi_2$ in (\ref{xi eq}).  

To find the shape of the star, we can replace $dp_y/dr$ in $\xi_0$ and $\xi_2$ with the TOV equation in (\ref{TOV}) and then take the limit $r \rightarrow R_*$.  This gives
\begin{equation} \label{xi0 p*0}
\begin{split}
\xi_0(R_*) &= \frac{R_*(R_*-2GM_*)}{GM_*}
\delta p_{*0}(R_*)
\\
\xi_2(R_*) &= \frac{R_*(R_*-2GM_*)}{GM_*}
\delta p_{*2}(R_*),
\end{split}
\end{equation}
where $\delta p_{*0}$ and $\delta p_{*2}$ are perturbations for the outermost fluid.  The shape of the star in the original coordinate system is then given by
\begin{equation} \label{shape}
\overline{R}_* = R_* + \xi_0(R_*) + \xi_2(R_*) P_2(\cos\theta),
\end{equation}
which follows from (\ref{xi def}).


\section{Numerical methods and example configurations}
\label{sec:numerical}

We solve the equations presented in the previous section numerically.  Solutions are identified by the central pressures $p_x(0)$ and the central values $\varpi_x(0) = \Omega_x  - \omega(0)$, which are used in (\ref{eta_x(0)}).  Ideally, we would be able to specify the angular velocities $\Omega_x$, but the angular velocities are determined for each solution from (\ref{u eta ext}).  Once the $p_x(0)$ and $\varpi_x(0)$ are specified, the system of ODEs can be integrated outward from $r = 0$. We describe our numerical procedure for solving the equations in the Appendix. 

Once we have a solution, the shape of the star can be computed using (\ref{xi0 p*0}) and (\ref{shape}).  The energy density and pressure curves are given by
\begin{equation}
\begin{split}
\mathcal{E}_x(r,\theta) &= \epsilon_x(r) + \Delta \mathcal{E}_x(r,\theta)
\\
\mathcal{P}_x(r,\theta) &= p_x(r) + \Delta \mathcal{P}_x(r,\theta),
\end{split}
\end{equation}
which can be computed using (\ref{delta eps P}), (\ref{delta script P R}), and (\ref{xi eq}).  The energy density and pressure can then be plotted in terms of the original radial coordinate $\bar{r}$ using (\ref{xi def}).

We now focus on the two-fluid system, where the first fluid describes the ordinary nuclear matter inside a neutron star and the second fluid describes dark matter.  We must choose an equation of state for each fluid.  For ordinary matter, we use SLy \cite{Douchin:2001sv, Haensel:2004nu}, which is a realistic equation of state.  Since precise properties of dark matter are largely unknown, we err on the side of simplicity and use a polytropic equation of state,
\begin{equation}
\mathcal{P}_\text{dm} = K \mathcal{E}_\text{dm}^\gamma,
\end{equation}
with $K = 100$ GeV$^{-4}$ and $\gamma = 2$.  We do not have reasons for choosing these equations of state beyond those mentioned and we consider these equations of state to be illustrative.

The slowly rotating approximation allows for different fluids to rotate at different angular velocities.  Since nongravitational interfluid interactions are negligible between ordinary matter and dark matter, we do not have have entrainment between the fluids \cite{Comer:2004yj}.  In the case of rapidly rotating ordinary matter initially residing in a cloud of nonrotating bosonic dark matter, it was found in \cite{DiGiovanni:2022mkn} that dark matter accretes onto the star and continues to be nonrotating.  It is therefore unclear what the relative angular velocities should be and, in this section and the next, we consider different possibilities.

\begin{figure*}
\centering
\includegraphics[width=6.5in]{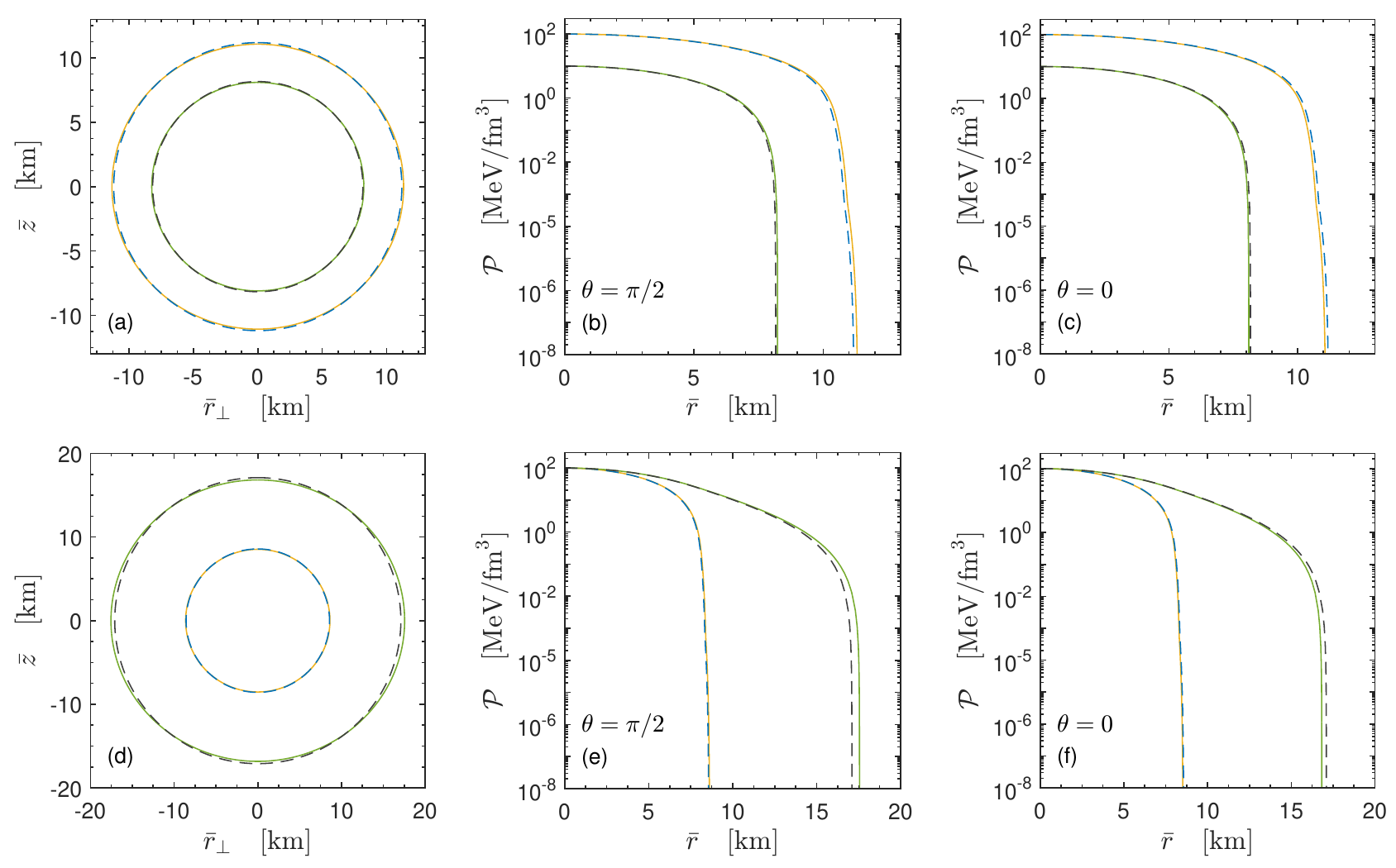}
\caption{Two example configurations are shown.  Properties for these configurations are listed in Table \ref{table1}.  In all plots, solid lines display the rotating solution, dashed lines display the nonrotating solution, blue/yellow is for ordinary matter with the SLy equation of state, and green/black is for dark matter with the polytropic equation of state.  (a--c) display a solution with a dark matter core, since ordinary matter extends beyond dark matter.  (d--f) display a different solution with a dark matter cloud, since dark matter extends beyond ordinary matter.  (a, d) plot a cross section in cylindrical coordinates.  (b, e) plot the pressure in the equatorial plane ($\theta = \pi/2$).  (c, f) plot the pressure along the rotation axis ($\theta = 0$).  }
\label{fig:1}
\end{figure*}

Figure \ref{fig:1} displays two example configurations.  In each plot, solid lines display the rotating solution, dashed lines display the nonrotating solution, blue/yellow is for ordinary matter with the SLy equation of state, and green/black is for dark matter with the polytropic equation of state.  The top row displays a configuration with a dark matter core and the bottom row displays a configuration with a dark matter cloud.  Figures \ref{fig:1}(a) and \ref{fig:1}(d) display a cross section in cylindrical coordinates, Figs.~\ref{fig:1}(b) and \ref{fig:1}(e) display the pressure in the equatorial plane (with $\theta = \pi/2$), and Figs.~\ref{fig:1}(c) and \ref{fig:1}(f) display the pressure along the rotation axis (with $\theta = 0$).  We can see the oblateness of the rotating solutions, with the fluids expanding in the equatorial plane and contracting along the rotation axis.  Properties of these solutions are listed in Table \ref{table1}.

\setlength{\tabcolsep}{5pt}
\begin{table} 
\normalsize
\begin{tabular}{l||c|c}
& Figs.~\ref{fig:1}(a)--\ref{fig:1}(c) & Figs.~\ref{fig:1}(d)--\ref{fig:1}(e) \\
\hline\hline
$p_\text{om}(0)$ \hfill [MeV/fm$^3$]  & $100$   & $100$ \\
$p_\text{dm}(0)$  \hfill [MeV/fm$^3$] & $10$    & $100$ \\  
$\varpi_\text{om}(0)$ \hfill [kHz] & 0.198 & 0.099 \\
$\varpi_\text{dm}(0)$ \hfill [kHz] & 0.077& 0.149 \\
$\tau_\text{om}$  \hfill [ms]                   & $2.76$  & $3.55$ \\
$\tau_\text{dm}$  \hfill [ms]                   & $4.13$  & $3.01$ \\
$\overline{R}_\text{om}^{(\theta=\pi/2)}$  \hfill [km]    & $11.1$ & $8.62$\\
$\overline{R}_\text{dm}^{(\theta=\pi/2)}$  \hfill [km]    & $8.24$ & $17.6$\\
$M_*$  \hfill [M$_\odot$]                       & $1.36 $ & $1.93$  \\
$\delta M$  \hfill [M$_\odot$]                  & $0.017$ & $0.031$ \\
$I_\text{om}/M_*^3$                             & $10.7$  & $0.918$ \\
$I_\text{dm}/M_*^3$                             & $0.238$ & $8.17$ \\
$Q(M_*/J^2)$                                    & $4.82$  & $4.57$
\end{tabular}
\caption{Various properties are listed for the two example configurations shown in Fig.~\ref{fig:1}.  $\tau_\text{om}$ and $\tau_\text{dm}$ are rotational periods.  The other quantities are defined in the main text.  The bottom three quantities are dimensionless.}
\label{table1}
\end{table}

The complete system of equations has scaling symmetries.  Specifically, given a solution, a new solution can be found by multiplying the first order perturbations by an arbitrary constant and the second order perturbations by the square of the same constant \cite{Hartle:1967he}.  For example, consider a single-fluid solution with angular velocity $\Omega$, angular momentum $J$, and quadruple moment $Q$.  We can quickly find a new solution such that the perturbations equal $\widetilde{u} = \alpha u$, $\widetilde{\eta} = \alpha {\eta}$, $\widetilde{v}_2 = \alpha^2 v_2$, $\widetilde{h}_2 = \alpha^2 h_2$, etc., where $\alpha$ is a constant.  From (\ref{u eta ext}) and (\ref{Q def}), the angular velocity, angular momentum, and quadruple moment for the new solution equal
\begin{equation}
\begin{split}
\widetilde{\Omega} = \alpha \Omega,
\qquad
\widetilde{J} = \alpha J,
\qquad
\widetilde{Q} = \alpha^2 Q.
\end{split}
\end{equation} 
By choosing $\alpha = \Omega_\text{new}/\Omega$, the new solution has angular velocity $\Omega_\text{new}$.

Our main focus is with the two-fluid case.  In the two-fluid case, we have two angular velocities and we can form a new solution with
\begin{equation}
\widetilde{\Omega}_1 = \alpha \Omega_1, 
\qquad
\widetilde{\Omega}_2 = \alpha \Omega_2,
\qquad
\widetilde{J} = \alpha J,
\qquad
\widetilde{Q} = \alpha^2 Q.
\end{equation}
The scaling symmetry allows us to quickly form solutions that preserve the ratio $\Omega_1/\Omega_2$.


\section{$I$-Love-$Q$}
\label{sec:ILoveQ}

Neutron stars rotate, which deforms the shape of the star.  The rotation can be characterized by the moment of inertia $I$ and the deformation can be characterized by the quadruple moment $Q$.  When a neutron star is in a binary orbit, the shape of the star is deformed by the gravitational field of the companion star.  This deformation can be characterized by the second Love number or equivalently by the tidal deformability $\lambda$.

The physics of neutron star interiors is contained within the equation of state.  Since we do not know what matter does at the extreme pressures found in the core of neutron stars, we do not know which equation of state we should be using.  This has led to a large number of proposed equations of state, each one based on different assumptions.  The celebrated $I$-Love-$Q$ relations \cite{Yagi:2013bca, Yagi:2013awa, Yagi:2016bkt} are relationships between the moment of inertia $I$, the tidal deformability $\lambda$, and the quadrupole moment $Q$ which are universal in that they are approximately independent of the equation of state.

Specifically, one computes the dimensionless quantities
\begin{equation} \label{dim I lamda Q}
\overline{I} \equiv \frac{I}{M^3_*},
\qquad
\overline{\lambda} \equiv \frac{\lambda}{M_*^5},
\qquad
\overline{Q} \equiv \frac{Q M_*}{J^2}.
\end{equation}
Using the scaling symmetries described in Sec.~\ref{sec:numerical}, we can see that for the single-fluid case, $\overline{I}$ and $\overline{Q}$ are independent of the angular velocity $\Omega$.  As a consequence, $\overline{I}$ and $\overline{Q}$ only depend on the central pressure $p(0)$.  This is also true for $\overline{\lambda}$ in the single-fluid case, which will be clear after we review the derivation of $\lambda$ below.  Plots of $\overline{I}$--$\overline{\lambda}$, $\overline{I}$--$\overline{Q}$, and $\overline{\lambda}$--$\overline{Q}$ are then one-parameter curves.  $I$-Love-$Q$ is the observation that these curves are approximately independent of the choice of equation of state and hence are universal \cite{Yagi:2013bca, Yagi:2013awa, Yagi:2016bkt}.

The study of dark matter admixed neutron stars has primarily focused on spherically symmetric properties, such as the mass and radius of the star.  Recently, Love numbers have been computed \cite{Das:2020ecp, Dengler:2021qcq, Collier:2022cpr}.  With fermionic dark matter, the only study of the moment of inertia that we are aware of, which requires only a first order analysis, is given in \cite{Leung:2012vea}.  As far as we are aware, the quadrupole moment has not previously been computed, which requires an analysis through second order.

The parameter space of nonrotating spherically symmetric dark matter admixed neutron stars can be described by the central pressures of the two fluids, $p_\text{om}(0)$ for ordinary matter and $p_\text{dm}(0)$ for dark matter.  If one of these central pressures is sufficiently large, the system is effectively a single-fluid system because the nondominant fluid has a negligible effect on bulk properties of the star \cite{Kain:2021hpk}.  The particular values of the central pressures at which this happens depend on the specific choices for the equations of state for each fluid.  If neither fluid dominates the system, the system is a truly mixed star and both fluids can affect bulk properties of the star.

We have found that this phenomenon continues to occur for slowly rotating dark matter admixed neutron stars.  This implies that if one of the central pressures is sufficiently large and the system is effectively a single-fluid system, the $I$-Love-$Q$ relations will be independent of the equations of state and angular velocities.  A question that we ask in this section is what happens to the $I$-Love-$Q$ relations when neither fluid dominates and we have a mixed star?

For dark matter admixed neutron stars, we continue to define the dimensionless $\overline{I}$, $\overline{\lambda}$, and $\overline{Q}$ as in (\ref{dim I lamda Q}), but we have a choice for $I$ and $J$:~$I$ and $J$ could be for a particular fluid or they could be the total moment of inertia and the total angular momentum.  Regardless of this choice, $\overline{I}$, $\overline{\lambda}$, and $\overline{Q}$ depend on the central pressures $p_\text{om}(0)$ and $p_\text{dm}(0)$ and $\overline{I}$ and $\overline{Q}$ depend on the ratio $\Omega_\text{om}/\Omega_\text{dm}$.


\subsection{Tidal deformability}

If a neutron star is in a binary orbit, the star will be tidally deformed by the gravitational field of the companion star.  This deformation is characterized by the second Love number or equivalently by the tidal deformability.  The tidal deformability can be measured through the gravitational wave signal of a binary inspiral \cite{LIGOScientific:2017vwq, LIGOScientific:2020aai}.  The tidal deformability for dark matter admixed neutron stars is computed in \cite{Nelson:2018xtr, Ellis:2018bkr, Karkevandi:2021ygv, Collier:2022cpr, Diedrichs:2023trk} for bosonic dark matter and in \cite{Das:2020ecp, Dengler:2021qcq, Collier:2022cpr} for fermionic dark matter.

The tidal deformability for a system with an arbitrary number of perfect fluids where the only interfluid interactions are gravitational is given by \cite{Char:2018grw, Nelson:2018xtr}
\begin{equation}
\lambda = \frac{2}{3} \kappa_2 R_*^5,
\end{equation}
where \cite{Hinderer:2007mb, Hinderer:2009ca}
\begin{align}
\kappa_2 &= \frac{8\mathcal{C}^5}{5}
(1 - 2\mathcal{C})^2 \left[2 + 2 \mathcal{C} (y_R - 1) - y_R \right]
\notag
\\
&\qquad \times \Bigl\{
2 \mathcal{C} [ 6 - 3 y_R + 3 \mathcal{C} (5y_R - 8)]
\notag
\\
&\qquad 
+ 4 \mathcal{C}^3 [13 - 11 y_R + \mathcal{C}(3 y_R - 2) + 2 \mathcal{C}^2 ( 1 + y_R) ]
\notag
\\
&\qquad 
+ 3(1 - 2 \mathcal{C})^2 [2 - y_R + 2\mathcal{C}(y_R - 1)] \ln(1 - 2\mathcal{C})
\Bigr\}^{-1}
\end{align}
is the second Love number and
\begin{equation}
y_R \equiv y(R_*),
\qquad
\mathcal{C} = M_*/R_*.
\end{equation}
$y(r)$ is found by solving 
\begin{equation} \label{y eq}
r \frac{d y(r)}{dr} + y^2(r) + y(r) F(r) + r^2 Q(r) = 0,
\end{equation}
where
\begin{equation}
\begin{split}
F &= \frac{r - 4\pi G r^3 \sum_x (\epsilon_x - p_x)}{r - 2 G M}
\\
Q &= \frac{4\pi G r}{r - 2 G M}
\biggl\{
\sum_x \left[
5\epsilon_x  + 9 p_x + \frac{\partial \epsilon_x}{\partial p_x} (\epsilon_x + p_x)
\right]
\\
&\qquad
- \frac{6}{4\pi G r^2} \biggr\}
- \left[ \frac{8\pi G r^3 \sum_x p_x + 2G M}{r (r - 2 GM r)}
\right]^2.
\end{split}
\end{equation}
Note that $y(r)$ depends only on nonrotating equilibrium quantities.  In practice, we solve Eq.~(\ref{y eq}) simultaneously with the system of equations presented in Sec.~\ref{sec:equations} using the inner boundary condition
\begin{equation}
\begin{split}
y(r) &= 2 
- r^2 \frac{4\pi G}{21} \sum_x
\biggl\{
33 p_x(0) + \epsilon_x(0) 
\\
&\qquad
+ 3 [\epsilon_x(0) + p_x(0)]
\frac{d\epsilon_x}{dp_x} \biggr|_{r=0} 
\biggr\} + O(r^4).
\end{split}
\end{equation}


\subsection{Results}

\begin{figure*}
\centering
\includegraphics[width=6.5in]{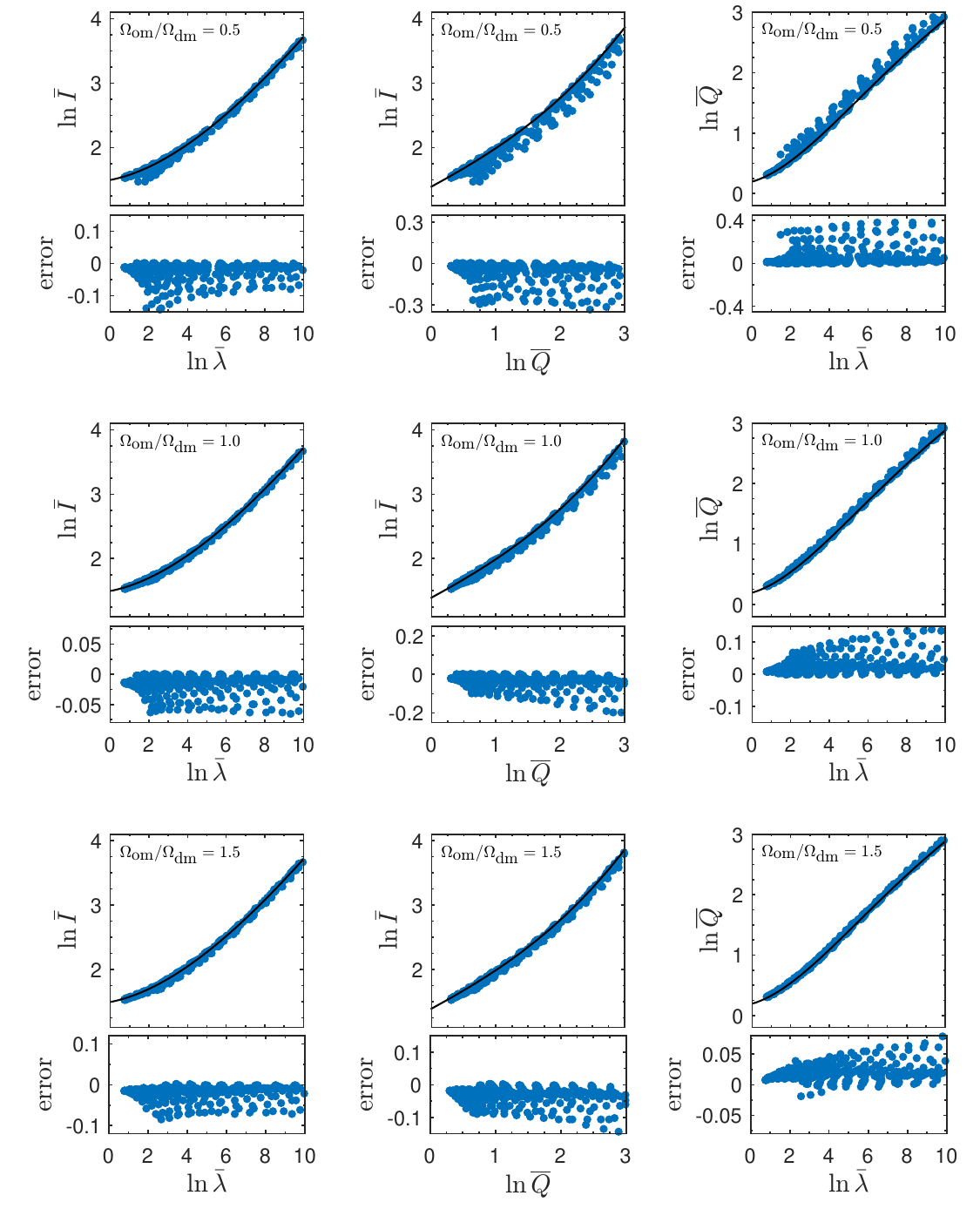}
\caption{$I$-Love-$Q$ plots using the definitions in (\ref{2f dim I lamda Q}).  The blue dots are our computed data values for $1 < p_\text{om}(0), p_\text{dm}(0)<10^3$ MeV/fm$^3$ and $\Omega_\text{om}/\Omega_\text{dm}$ as indicated in the figures.  The black curves are the single-fluid fitting curves given in (\ref{fitting curve}).  The bottom panel in each figure gives the relative error defined by (\ref{error}).  For a (single-fluid) neutron star, the relative error is less than 1\% \cite{Yagi:2013bca, Yagi:2013awa, Yagi:2016bkt}.  This figure indicates that dark matter admixed neutron stars can have a relative error greater than 1\%.}
\label{fig:2}
\end{figure*}

We begin by straightforwardly generalizing the standard single-fluid $I$-Love-$Q$ plots to two fluids.  Specifically, we generalize the dimensionless quantities in (\ref{dim I lamda Q}) to
\begin{equation} \label{2f dim I lamda Q}
\overline{I} \equiv \frac{I_\text{om} + I_\text{dm}}{M^3_*},
\qquad
\overline{\lambda} \equiv \frac{\lambda}{M_*^5},
\qquad
\overline{Q} \equiv \frac{Q M_*}{(J_\text{om} + J_\text{dm})^2}.
\end{equation}
These quantities depend on $p_\text{om}(0)$, $p_\text{dm}(0)$, and $\Omega_\text{om}/\Omega_\text{dm}$.  We show results in Fig.~\ref{fig:2} for $1 < p_\text{om}(0), p_\text{dm}(0)<10^3$ MeV/fm$^3$ and a few representative values of $\Omega_\text{om}/\Omega_\text{dm}$. To construct these plots, for each value of $p_\text{om}(0)$ and $p_\text{dm}(0)$, we fix $\varpi_\text{om}(0)$ to an arbitrary value and search through values for $\varpi_\text{dm}(0)$ until the resulting solution has the required ratio $\Omega_\text{om}/\Omega_\text{dm}$.

In each plot, the blue dots are our computed data values and the black curves are the single-fluid fitting curves given by
\begin{equation} \label{fitting curve}
\ln y_i = a_i + b_i \ln x_i + c_i (\ln x_i)^2 + d_i (\ln x_i)^3 + e_i (\ln x_i)^4,
\end{equation}
where the $x_i,y_i$ are the $\overline{I}, \overline{\lambda}, \overline{Q}$ plotted on the horizontal and vertical axes and the values of the fitting coefficients $a_i, b_i, c_i, d_i, e_i$ can be found in \cite{Yagi:2013awa}.  In the bottom panel of each plot, the relative error is defined by
\begin{equation} \label{error}
\text{error} = \frac{y^\text{data}_i - y_i}{y_i}.
\end{equation}

In the single-fluid case, the relative error is less than 1\% \cite{Yagi:2013bca, Yagi:2013awa, Yagi:2016bkt}.   An immediate feature of Fig.~\ref{fig:2} is that the relative error can be greater than 1\% for dark matter admixed neutron stars.  This occurs when the star is truly mixed and neither fluid is dominating.  This is the case for all three types of plots and for all three angular velocity ratios.  We note that our results are consistent with those in \cite{Yeung:2021wvt}.

The Neutron Star Interior Composition Explorer (NICER) has recently made precise measurements of the mass and radius of millisecond pulsars \cite{Riley:2019yda, Miller:2019cac, Riley:2021pdl, Miller:2021qha}.  NICER measures x-ray emissions from the surface of the star.  These electromagnetic observations can be used to determine the moment of inertia and angular momentum \cite{Lattimer:2004nj, Silva:2020acr}.  Since these observations are electromagnetic, they direly probe ordinary matter and not dark matter.  The two-fluid generalization we made in Eq.~(\ref{2f dim I lamda Q}) and in Fig.~\ref{fig:2} does not adequately take this into account.  Assuming measurements for $I_\text{om}$ and $J_\text{om}$, we should define the dimensionless quantities as
\begin{equation} \label{om dim I lamda Q}
\overline{I}_\text{om} \equiv \frac{I_\text{om}}{M^3_*},
\qquad
\overline{\lambda} \equiv \frac{\lambda}{M_*^5},
\qquad
\overline{Q}_\text{om} \equiv \frac{Q M_*}{J_\text{om}^2}.
\end{equation}

\begin{figure*}
\centering
\includegraphics[width=6.5in]{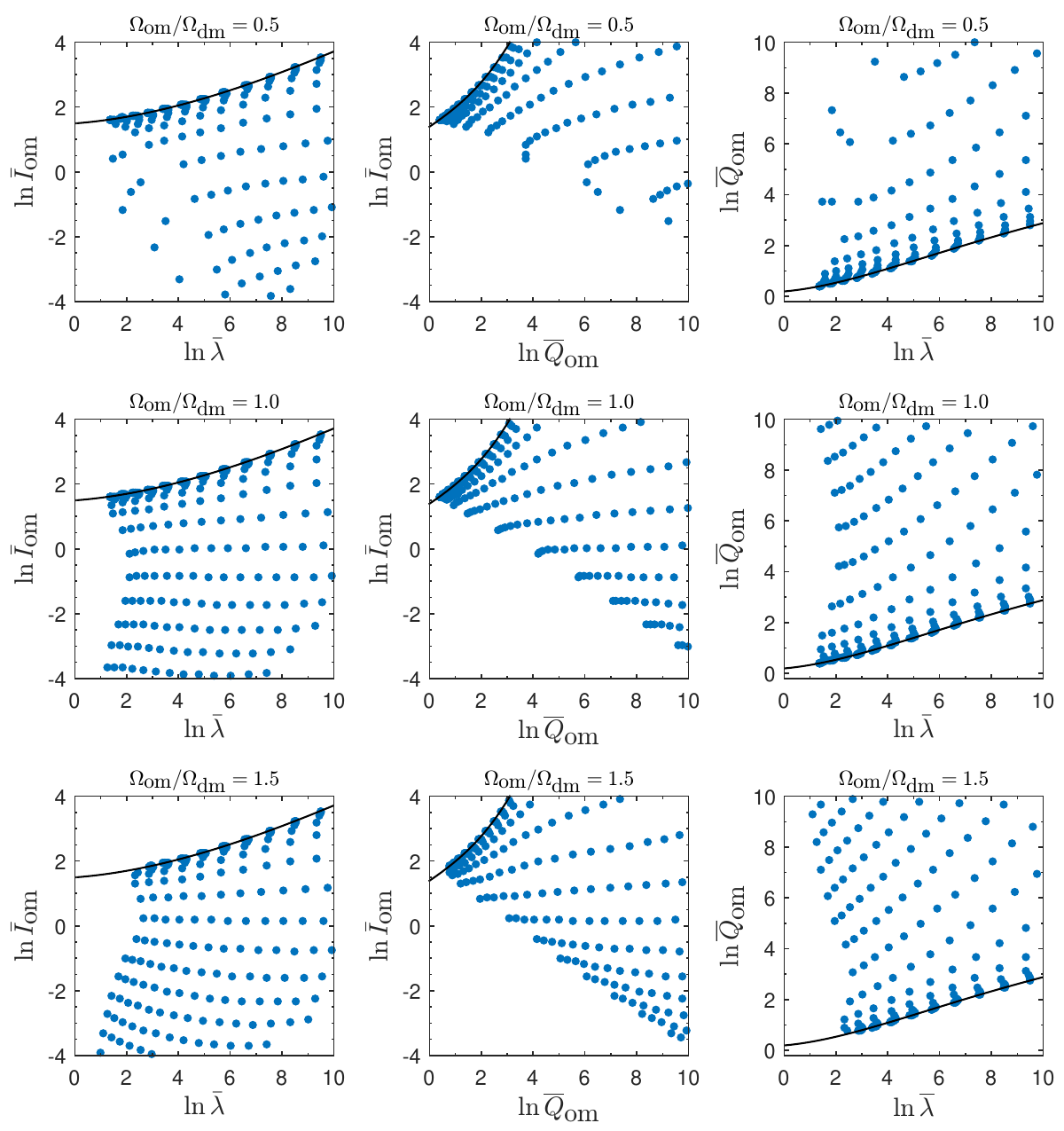}
\caption{This figure is analogous to Fig.~\ref{fig:2}, except it uses the dimensionless variables defined in (\ref{om dim I lamda Q}).  The dimensionless variables in (\ref{om dim I lamda Q}) assume electromagnetic measurements for the moment of inertia and angular momentum which probe ordinary matter alone and not dark matter.  This figure shows large deviations from the single-fluid fitting curves.}
\label{fig:3}
\end{figure*}

We show results in Fig.~\ref{fig:3} using the definitions in (\ref{om dim I lamda Q}) and for when dark matter does not dominate the star.  Comparing with Fig.~\ref{fig:2}, we can immediately see large deviations from the single-fluid fitting curves.  Indeed, the deviations are so large that we do not include panels for the relative error.  Note also that the results in Fig.~\ref{fig:3} cannot be parametrized by a one-parameter curve.  An important takeaway from Fig.~\ref{fig:3} is that if measurements of $\overline{I}$, $\bar{\lambda}$, $\overline{Q}$ deviate significantly from the single-fluid $I$-Love-$Q$ curves and if such measurements for multiple stars do not lie along a curve, then we have strong evidence for the existence of dark matter admixed neutron stars.


\section{Conclusion}
\label{sec:conclusion}

Using Hartle's slowly rotating approximation, we derived equations describing a rotating system with an arbitrary number of perfect fluids with only gravitational interfluid interactions.  We then specialized to the two-fluid case for describing rotating dark matter admixed neutron stars with fermionic dark matter. The two-fluid case is equivalent to the formalism developed by Andersson and Comer \cite{Andersson:2000hy} in the limit that inter-fluid interactions are neglected.

Using our two-fluid model, we studied $I$-Love-$Q$ relations.  For a standard two-fluid system, we found deviations from the single-fluid results that are consistent with previous results \cite{Yeung:2021wvt, Aranguren2}.  However, our expectation is that measurements of the moment of inertia and the angular momentum will be in terms of ordinary matter alone and will not directly probe dark matter.  When we parametrize the $I$-Love-$Q$ dimensionless variables in terms of the moment of inertia and the angular momentum for ordinary matter, we found significant deviation from the single-fluid relations and that $I$-Love-$Q$ cannot be described by one-parameter fitting curves.  Our results immediately suggest a method for determining if neutron stars contain sufficient quantities of dark matter such that dark matter can affect bulk properties of the star.

We note that our analysis has only made use of one choice for the equation of state for ordinary matter and one choice for the equation of state for dark matter.  Further, we have only considered a few choices for the ratio of angular velocities.  Our aim with this paper is to show that dark matter admixed neutron stars can deviate significantly from the standard result and not to make a broad study of different equations of state, as has been done for single-fluid $I$-Love-$Q$.  However, we have considered other equations of state and found results qualitatively similar to those presented in this paper. Nevertheless, a systematic analysis using various equations of state would be interesting.


\appendix

\section{Numerical procedure}
\label{appdendix}

In this appendix, we outline our numerical procedure for solving the various equations presented in Sec.~\ref{sec:equations}. Solutions are identified by the central pressures $p_x(0)$ and the central values $\varpi_x(0)$. Once these values are specified, the system of ODEs can be integrated outward from $r = 0$.

The first step is to integrate the TOV equations in (\ref{TOV}) using an arbitrary value for $\nu(0)$ (in practice, we use $\nu(0) = 0$) to find the equilibrium solution. $R_*$ is defined as the radial position where the pressure of the outermost fluid drops to zero. From (\ref{M* def}), $M_* = M(R_*)$. A look at the TOV equations shows that $\nu(r)$ can be shifted by a constant with the result still being a solution.  We can therefore shift $\nu(r)$ such that the boundary condition in (\ref{nu ext}) is satisfied.  Upon making the shift, the updated inner boundary value is
\begin{equation} \label{new nu}
\nu (0) \rightarrow  \nu(0) - \left[ \nu(R_*) - \ln \left(1 - \frac{2GM_*}{R_*}\right)\right].
\end{equation}
From this point forward, we exclusively use this updated value for $\nu(0)$. If we were to integrate the TOV equations again, but now using (\ref{new nu}) as the inner boundary condition for $\nu(0)$, we would obtain the complete equilibrium solution.

The next step is to integrate the full system of equations outward from $r = 0$ to $r = R_*$. This system includes the TOV equations in (\ref{TOV}), along with the bottom equation in (\ref{chi u eqs}) for $\eta_x$, (\ref{dux dr}) for $u_x$, (\ref{m0 deltapx0 eqs}) for $m_0$ and $\delta p_{x0}$, (\ref{v2 h2 eqs}) for $v_2$ and $h_2$, and (\ref{y eq}) for $y$. At the end of the integration, we use the results to solve Eq.~(\ref{v2 h2 exterior}) for the constants $C$ and $K$ and Eqs.~(\ref{integrals of motion}) and (\ref{m2 eq}) to solve for $\delta p_{x2}$ and $m_2$.  We now have the complete interior solution and can straightforwardly compute properties of the star, such as the angular velocities $\Omega_x$ using (\ref{u eta ext}), the total moment of inertia $I$ using (\ref{Ix def}), the mass correction $\delta M$ using (\ref{delta M}), and the quadrupole moment $Q$ using (\ref{Q def}).


\acknowledgments

J.~C.~was funded by an Alden Fellowship.  X.~Z.~was funded by the Weiss Summer Research Program.




%

\end{document}